\documentclass[a4paper, 10pt, oneside, onecolumn, preprint]{elsarticle}
\usepackage{graphicx}
\usepackage{subfig}

\usepackage{graphicx}% Include figure files
\usepackage{dcolumn}% Align table columns on decimal point
\usepackage{bm}% bold math

\begin{document}

\title{Ordering in a fibrinogen layer}

\author[uj]{Michal Ciesla}
\ead{michal.ciesla@uj.edu.pl}
\author[uj,pan]{Jakub Barbasz}
\ead{ncbarbas@cyf-kr.edu.pl}

\address[uj]{M. Smoluchowski Institute of Physics, Jagiellonian University, 30-059 Kraków, Reymonta 4, Poland.}
\address[pan]{Institute of Catalysis and Surface Chemistry, Polish Academy of Sciences, 30-239 Kraków, Niezapominajek 8, Poland.}

\begin{abstract}
Ordered protein layers are the subject of active biomedical research for their usually  interesting physicochemical properties, e.g. permeability, stiffness and pours structure. In presented work, we focused on layers build of fibrinogen molecules characterised by strong shape anisotropy. Using Random Sequential Adsorption (RSA) method, we simulated adsorption process in which the orientation of adsorbate was described by a non-uniform probability distribution. Thus obtained covering layers had different level of global orientational ordering. This allowed us to find dependence between main properties of layers, such as maximal random coverage ratio, and order parameter. For better description and deeper understanding of obtained structures, the autocorrelation function as well as distribution of uncovered space were determined. Additionally, we calculated the Available Surface Function (ASF), which is essential for determining adsorption kinetics.
\end{abstract}

\maketitle
\section{Introduction}
Structure of monolayers formed during adsorption process is strongly related to properties of underlying substrate, environmental conditions and adsorbate molecules themselves. Earlier works on structured protein coverages indicate a high potential of such layers for material, medical and food sciences as well as pharmaceuticals and cosmetics industries. In the present study we assumed that strongly anisotropic fibrinogen molecules tend to align along specific axis. Experiments performed earlier show that ordering can have considerable impact on covering layers, e.g. \cite{bib:Keller2011, bib:Zemla2012}. Our aim is to find out how the ordering affect maximal random coverage ratio, coverage structure, and adsorption kinetics. To achieve that, we performed extensive numerical simulations on fibrinogen adsorption using RSA algorithm.
\section{Model}
\begin{figure}[htb]
\centerline{%
\includegraphics[width=6cm]{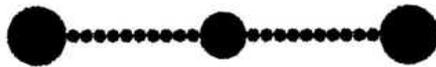}}
\caption{Approximation of fibrinogen molecule shape used in simulations. The side balls have diameter of $6.7 nm$ whereas the middle one is a little smaller $5.3 nm$. Each of the small spheres between them is of $1.5 nm$ diameter.}
\label{fig:fibrynogenmodel}
\end{figure}

Fibrinogen molecule is modeled as a component of three spheres of $6.7$, $5.3$ and $6.7 nm$ in diameter, connected by two chains of ten small $1.5 nm$ spheres; see Fig.\ref{fig:fibrynogenmodel}.
Such a model was used earlier in \cite{bib:Adamczyk2010, bib:Adamczyk2011} and it turned out to be the most effective in reproducing maximal coverages and adsorption kinetics obtained in experiments.
\par
\begin{figure}[htb]
\vspace{1cm}
\centerline{%
\includegraphics[width=6cm]{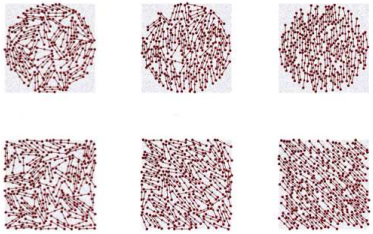}}
\caption{Example coverages for two different collector shapes and three different $\sigma^2$. From left to right $\sigma^2=1.0, 0.5$ and $0.1$ which according to (\ref{eq:order}) corresponds to order parameter $q=0.188, 0.668$ and $0.982$ respectively. The expected direction was $\bar{\varphi}=\pi/2$ for round collectors and $\pi/4$ for square ones.}
\label{fig:examples}
\end{figure}

 
Fibrinogen molecules are placed on a ﬂat homogeneous collector surface according to Random Sequential Adsorption (RSA) algorithm \cite{bib:Feder1980}, which iteratively repeats the following steps:
\begin{description}
\item[1.] a virtual fibrinogen molecule is created with its centre position on a collector is chosen randomly according to the uniform probability distribution; 
\item[2.] the molecule orientation is chosen randomly according to normal probability distribution having specific expected value $\bar{\varphi}$ and variance $\sigma^2$;
\item[3.] an overlapping test is performed for previously adsorbed nearest neighbours of the virtual molecule. The test checks if surface-to-surface distance between each of the spheres is greater than zero;
\item[4.] if there is no overlap the virtual molecule is irreversibly adsorbed and added to an existing covering layer. Its position does not change during further calculations;
\item[5.] if there is an overlap the virtual fibrinogen molecule is removed and abandoned.
\end{description}
The number of RSA iterations $N$ is typically expressed in dimensionless time units:
\begin{equation}
t = N\frac{S_F}{S_C},
\end{equation}
where $S_F = 127.918 nm^2$ is an area covered by single fibrinogen molecule and $S_C$ is a collector size. In case of our simulations algorithm was stopped after $t = 10^4$.
Example coverages obtained this way are presented in Fig.\ref{fig:examples}.
To check the importance of boundaries shape, simulations were performed using both round and square collectors. Obtained results do not show any significant dependence on collector shape. For further analysis, coverages generated using round collector of $500 nm$ diameter were chosen.
\par
As mentioned above, the orientational order inside a layer is controlled by a parameter $\sigma^2$ bound up with the width of Gaussian probability density function used during simulations. However in possible experiments, ordering can be evoked by many different factors, e.g. collector structure, electrostatic interactions or flows. Each of them will generally produce different orientation probability distributions with variance being typically a nonlinear function of experimentally controlled parameters. On the other hand, this study focuses on properties of ordered layers regardless of how they were created. Therefore, it is worth to introduce separate parameter describing orientational order, irrespectively of underlying phenomenons promoting parallel alignment of molecules. The order parameter can be expressed as follows \cite{bib:Ciesla2012a}:
\begin{equation}
q = 2\left[ \frac{1}{N}\sum_{i=1}^n \left(x_i \cos \phi + y_i \sin \phi \right)^2 - \frac{1}{2} \right],
\label{eq:order}
\end{equation}
where $n$ is a number of molecules in a layer, $[x_i, y_i]$ describes a unit vector parallel to orientation of $i$-th fibrinogen, and $\phi$ denotes a mean direction. Parameter $q$ defined as above is $0$ for totally disordered layer and rises up to $1$ with growing orientational order. The maximum value is reached when all molecules inside a coverage are parallely aligned.
\section{Results and discussion}
\subsection{Coverage ratio}
\begin{figure}[htb]
\vspace{1cm}
\centerline{%
\includegraphics[width=6cm]{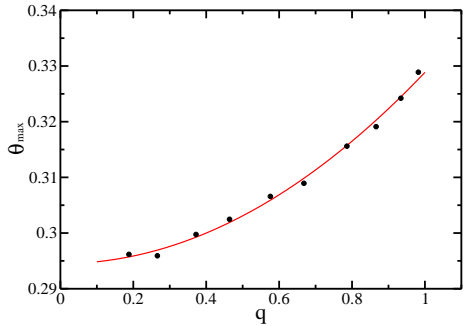}}
\caption{The maximal random coverages dependence on the order parameter  (\ref{eq:order}). Dots are simulation data whereas the solid line corresponds to square fit:
$\theta_{max} = 0.034\cdot q^2 + 0.295$. }
\label{fig:theta_q}
\end{figure}

The maximal random coverage $\theta_{max}$ was estimated by measuring the number of adsorbed molecules $n$:
\begin{equation}
\theta_{max} = n\frac{S_F}{S_C}.
\end{equation}
Determining its dependence on orientational order was the main aim of this paper. Values obtained from simulations were averaged over at least $20$ different collectors. Results are presented in Fig.\ref{fig:theta_q}.
As expected the maximal random coverage grows with orientational order, although the overall impact is only 10\%. The trend of the growth can be successfully fitted with a parabola in which the free term is in concordance with the maximal random coverage for fibrinogen layers determined in \cite{bib:Adamczyk2010}.  
\subsection{Autocorrelations and structure}
\begin{figure}[htb]
\vspace{1cm}
\centerline{%
\includegraphics[width=6cm]{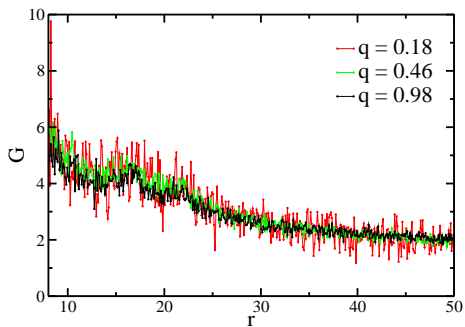}}
\caption{The density autocorrelations for coverages characterised by three different order parameters.}
\label{fig:g}
\end{figure}

To study the structure of obtained fibrinogen layers in more detail we analysed density autocorrelations and distribution of uncovered places.The density autocorrelation function is defined here as:
\begin{equation}
G(r) = \frac{P(r)}{2\pi r \rho},
\end{equation}
where $P(r)$ is a probability of finding two molecules in distance $r$ and $\rho$ is mean density of molecules inside a covering layer. Fig.\ref{fig:g} shows $G(r)$ function graph for three different values of the order parameter.
It is noteworthy that in case of fibrinogen lack of global order does not mean lack of order at all. It can be noticed even on snapshots presented in fig.\ref{fig:examples}; however, presented $G(r)$ proves that. Apart from the statistical noise, all plots look similar. In all the cases local maxima and minima are in the same positions and correspond only to geometrical properties of the fibrinogen molecule.
\par
\begin{figure}[htb]
\vspace{1cm}
\centerline{%
\includegraphics[width=6cm]{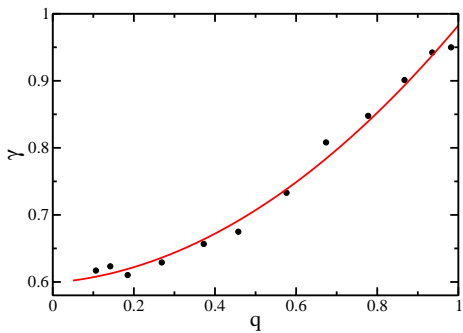}}
\caption{The decay constant $\gamma$ defined by (\ref{gamma}) dependence on orientational ordering. Dots are simulation data whereas the solid line is a quadratic fit: $\gamma = 0.335\cdot q^2 + 0.049 \cdot q + 0.599$.}
\label{fig:g_o}
\end{figure}

Another interesting characteristic of a coverage structure is distribution of uncovered places. It can be of particular importance when the layer undergoes further adsorption of different molecules. To measure this property, the uncovered places were filled with balls of a different size giving the higher priority to larger spheres. It occurs that the size distribution of balls for all values of the order parameter $q$ is well approximated by an exponential law:
\begin{equation}
N(r) = A \exp(-\gamma r)
\label{gamma}
\end{equation}
where $N(r)$ is the number of balls having radius $r$ and $A$ is a normalisation constant.
The decay constant $\gamma$ depends on orientational order inside a layer as shown in Fig.\ref{fig:g_o}.
Lower values of $\gamma$ for disordered layers mean more large uncovered spots, which is in line with the expectations.
\subsection{Kinetics}
\begin{figure}[htb]
\vspace{1cm}
\centerline{%
\includegraphics[width=6cm]{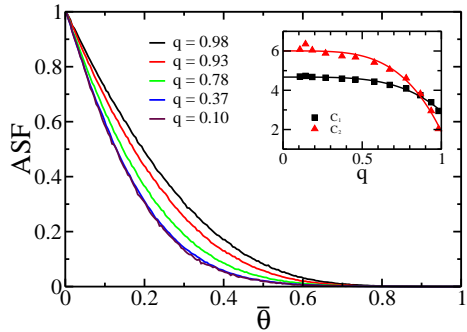}}
\caption{Available Surface Function for different order parameter $q$. Inset shows coefficients $C_1$ and $C_2$ used in low coverage approximation (\ref{asffit}) and their dependence on $q$. Squares and triangles represents data obtained from simulations whereas lines are exponential fits: $C_1(q) = 4.709 - 0.015 \cdot \exp(4.799 q)$ and $C_2(q) = 6.158 - 0.053 \cdot \exp(4.412 q)$.}
\label{fig:asf}
\end{figure}

There are two factors affecting the adsorption kinetics measured in experiments. The first one is transport process shifting molecules to a surface proximity. It depends on a given experiment's environmental conditions and therefore hardly enters the general theoretical analysis. The second factor is varying probability of adsorption. It decreases in time due to diminishing area of uncovered collector surface. The dependence between adsorption probability and temporary coverage ratio is commonly known as Available Surface Function (ASF) and it can be easily measured in RSA simulations. Results are shown in Fig.\ref{fig:asf}.
To compare data for different $q$, the coverage ratios were normalised: $\bar{\theta} = \theta / \theta_{max}$. At the limit of low coverage, the $ASF(\bar{\theta})$ is commonly approximated by a quadratic fit \cite{bib:AdamczykBook}:
\begin{equation}
ASF(\bar{\theta}) = 1 - C_1 \bar{\theta} + C_2 \bar{\theta}^2.
\label{asffit}
\end{equation}
Coefficients $C_1$ and $C_2$ can be calculated using the least squares method. Their values for different order parameter $q$ are presented on the inset in Fig.\ref{fig:asf}. Both plots in this figure shows that the ASF decays slower for highly ordered layers which is in agreement with the expectations and previous results for maximal coverage ratio.
\par
\begin{figure}[htb]
\vspace{1cm}
\centerline{%
\includegraphics[width=6cm]{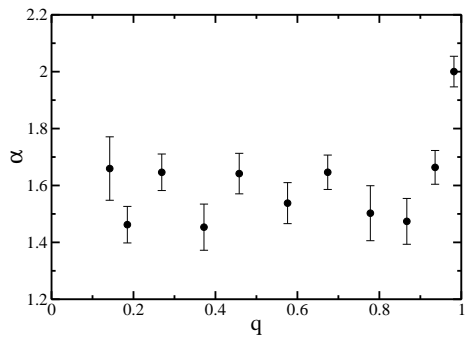}}
\caption{Dependence between $\alpha$ and orientational order $q$.}
\label{fig:a}
\end{figure}

At the jamming limit, adsorption practically does not depend on transport process because its speed is restricted by amount of uncovered spots. Therefore, due to it's universality, this case has been extensively studied. For example, for spherical molecules it was shown analytically that the coverage's growth obeys the so called Feder's law \cite{bib:Swendsen1981, bib:Privman1991}:
\begin{equation}
\theta_{max} - \theta(t) \sim t^{-1/d},
\label{fl}
\end{equation}
where $d$ is a dimension of a space (collector and adsorbate). Relation (\ref{fl}) has been proved numerically for a number of dimensions \cite{bib:Torquato2006, bib:Ciesla2012b} and different adsorbates \cite{bib:Ciesla2012a}. Since $ASF(\theta) \sim d\theta / dt$, equation (\ref{fl}) is equivalent to:
\begin{equation}
ASF(\bar{\theta}) \sim \left( 1 - \bar{\theta} \right)^\alpha,
\label{asffit}
\end{equation}
where $\alpha = d+1$. The $\alpha$ values can be obtained from plot \ref{fig:asf} usind the least squares approximation method. Relation (\ref{asffit}) is valid within $\theta \to \theta_{max}$ limit, therefore only data for $\bar{\theta} >0.8$ were used for estimation. Results are shown in Fig.\ref{fig:a}
The statistical errors are quite large due to poor statistics; however the correlation coefficient in all cases exceeded $0.9$ which means quite good agreement with (\ref{asffit}). As $\alpha<3$ the Feder's law is never valid for fibrinogen adsorption in 2D. Parameter $\alpha$ fluctuates around $1.5$ to rise up to $2.0$ for the most ordered layer. It suggests that ordered fibrinogen adsorption kinetics is similar to one dimensional adsorption. Smaller values of $\alpha$ for disordered phase are unexpected and do not agree with commonly used fit for elongated molecules: $ASF(\bar{\theta}) = (1 + a_1 \bar{\theta} + a_2 \bar{\theta}^2 + a_3 \bar{\theta}^3)(1-\bar{\theta})^4$ \cite{bib:Ricci1992}.
\section{Summary}
The global orientational ordering inside a fibrinogen layer affects the maximal random coverage ratio, however the overall increase is smaller than 10\%. The density autocorrelation does not change as the local orientational ordering is present also for globally disordered layers. The analysis of remaining uncovered spots shows that larger areas are more common in disordered coverages. Adsorption kinetics depends on adsorbate alignment and is faster for parallely oriented molecules. RSA of fibrinogen does not obey the Feder's law in its original form. For ordered layers the reaction speed is similar to one dimensional systems.
\par
This work was supported by grant MNiSW/0013/H03/2010/70.

\end{document}